\documentclass[preprint2]{aastex}

\shorttitle{Unveiling of SU Aurigae } \shortauthors{Chakraborty et al.}

\begin{document}

%% LaTeX will automatically break titles if they run longer than
%% one line. However, you may use \\ to force a line break if
%% you desire.

\title{Unveiling Su Aurigae in the near Infrared:\\
New high spatial resolution results using Adaptive Optics}

%% Use \author, \affil, and the \and command to format
%% author and affiliation information.
%% Note that \email has replaced the old \authoremail command
%% from AASTeX v4.0. You can use \email to mark an email address
%% anywhere in the paper, not just in the front matter.
%% As in the title, you can use \\ to force line breaks.

\author{Abhijit Chakraborty, \& Jian Ge}
\affil{Department of Astronomy \& Astrophysics, 525 Davey
Laboratory, Pennsylvania State University, University Park, PA
16802} \email{abhijit@astro.psu.edu} \email{jian@astro.psu.edu}

%% Notice that each of these authors has alternate affiliations, which
%% are identified by the \altaffilmark after each name.  Specify alternate
%% affiliation information with \altaffiltext, with one command per each
%% affiliation.

%\altaffiltext{1}{Visiting Astronomer, Cerro Tololo Inter-AmericanObservatory.
%CTIO is operated by AURA, Inc.\ under contract to the National Science
%Foundation.}
%\altaffiltext{2}{Society of Fellows, Harvard University.}
%\altaffiltext{3}{present address: Center for Astrophysics,
%    60 Garden Street, Cambridge, MA 02138}
%\altaffiltext{4}{Visiting Programmer, Space Telescope Science Institute}
%\altaffiltext{5}{Patron, Alonso's Bar and Grill}

%% Mark off your abstract in the ``abstract'' environment. In the manuscript
%% style, abstract will output a Received/Accepted line after the
%% title and affiliation information. No date will appear since the author
%% does not have this information. The dates will be filled in by the
%% editorial office after submission.

\begin{abstract}
We present here new results on circumstellar nebulosity around SU
Aurigae, a T-Tauri star of about 2 solar mass and 5 Myrs old at
152 pc in the J, H and K bands using high resolution adaptive
optics imaging (0$\farcs$30) with the Penn state IR Imaging Spectrograph 
(PIRIS) at the 100 inch Mt. Wilson telescope.

A comparison with HST STIS optical (0.2 to 1.1 micron) images
shows that the orientation of the circumstellar nebulosity in the
near-IR extends from PAs 210 to 270 degrees in H and K bands and
up to 300 degrees in the J band. We call the circumstellar
nebulosity seen between 210 to 270 degrees as 'IR nebulosity'. We
find that the IR nebulosity (which extends up to 3.5 arcsecs in J
band and 2.5 arcsecs in the K band) is due to scattered light from
the central star. The IR nebulosity is either a cavity formed by
the stellar outflows or part of the circumstellar disk. We present
a schematic 3-dimensional geometrical model of the disk and jet of
SU Aur based on STIS and our near-IR observations. According to
this model the IR nebulosity is a part of the circumstellar disk
seen at high inclination angles. The extension of the IR
nebulosity is consistent with estimates of the disk diameter of 50
to 400 AU in radius, from earlier mm, K band interferometric
observations and SED fittings.
\end{abstract}

%% Keywords should appear after the \end{abstract} command. The uncommented
%% example has been keyed in ApJ style. See the instructions to authors
%% for the journal to which you are submitting your paper to determine
%% what keyword punctuation is appropriate.

\keywords{stars: individual (SU Aurigae) - stars: pre-main-sequence - stars:
formation - scattering
-techniques: high angular resolution - techniques: image processing}

%% From the front matter, we move on to the body of the paper.
%% In the first two sections, notice the use of the natbib \citep
%% and \citet commands to identify citations.  The citations are
%% tied to the reference list via symbolic KEYs. The KEY corresponds
%% to the KEY in the \bibitem in the reference list below. We have
%% chosen the first three characters of the first author's name plus
%% the last two numeral of the year of publication as our KEY for
%% each reference.

\section{Introduction}
SU Aurigae (SU Aur) is a T-Tauri star located in the
Taurus-Aurigae complex of dark molecular clouds at a distance of
152 pc (de Warf et al. 2003, Hipparcos catalog). Its spectral type
is G2III, mass $\sim$ 2$M_\odot$ and age is about 4 to 5 million
years (de Warf et al. 1998, de Warf et al. 2003). Recent
observations by Nadalin et al. (2000) have shown short time
variability in the B band magnitude that they attribute to
proto-planetary materials orbiting in the circumstellar disk.
Petrov et al. (1996) have found existence of a gas outflow from
the star from their spectroscopic results.

The HST/STIS coronagraphic observations by Grady et al. (2002)
from 0.2 to 1.0 microns have revealed fan like structures
extending up to a distance of 12-15 arcsecs in the west to
south-west direction. These, according to the authors, are mainly
reflection nebulae scattering the light of the central star. In
addition they also detected streaming filamentary structures going
radially outwards from the star which could be due to gas outflow
either from the star or from the parent molecular cloud.

Our motivation comes from the STIS images of SU Aur, and from the
fact that no report of high spatial resolution near-infrared
images of the star could be found in the literature to this date.
Therefore, we observed the source in the near-infrared (J, H and K
band images) using adaptive optics and applied the technique of
PSF subtraction to investigate the circumstellar region of SU Aur
with high spatial resolution (0$\farcs$25 to 0$\farcs$30). In this 
paper we present the results of our investigation (of SU Aur). 
Section 2 describes observations and data analysis procedure, section 
3 introduces the results, in section 4 we discuss our results and 
present our conclusions in section 5.

\section{Observations and Data Analysis}
\subsection{Observations}
SU Aurigae was observed in the J,H and K photometric bands during
October 2002 at the 100 inch telescope of Mt.Wilson using its
natural star Adaptive Optics (AO) system (Shelton et al. 1995) and
the Penn state near-IR Imager and Spectrograph (PIRIS, Ge et al.
2003). The detector is a 256$\times$256 PICNIC array with 40
micron pixels. The plate scale is 0$\farcs$082 per pixel providing
a field of view of 21$\farcs$0. PIRIS is also equipped with cold
pupil masks in the pupil plane to reduce the thermal background
(particularly in the K band).

The night of observations (19 Oct. 2002) was photometric with
sub-arcsec seeing and the FWHM of star's Point Spread Function
(PSF) after the AO correction was measured to be 0$\farcs$25.
Seven images of SU Aur with 60 second exposures in each
photometric band were recorded when the object was near zenith
(air mass close to 1.00). Two PSF stars were also observed namely
BD +40 248 (Spectral type = G2V, PSF1) and Gl 46.1 (Spectral type
= G5V, PSF2). Gl 46.1 was observed 3.5 hours before the SU Aur
observations with similar air mass (or close to the zenith) while
the other PSF star BD +40 248 was observed immediately before the
SU Aur observations but at a higher air mass (1.6). This allowed
us to determine the stability of the PSF during the night of
observations as the telescope tracks different sources over
different air masses. We find that the FWHMs of the PSF stars were
0$\farcs$25 and 0$\farcs$29 respectively, and the overall shape
did not change by more than 5\% between the two PSF stars. In fact
another star Gl 230 (of spectral type G2) was observed (at air
mass=1.1) as a part of a different program immediately after the
SU Aur observations by the same instrument and its PSF has a FWHM
of 0$\farcs$28 and has similar shape.

The PSF stars were selected based on their similar brightness and
spectral type to SU Aurigae. For instance, the J, H, K magnitudes
of BD +40 248 are 5.76, 5.35, 5.27 and Gl 46.1 are 5.50, 4.96,
4.75 respectively (from 2MASS All Sky Point Source Catalog). From
the same catalog the SU Aur J, H, K magnitudes are 7.20, 6.56 and
5.99 respectively. Three exposures of 1, 10, 20 and 30 seconds in
each photometric bands (J, H and K) of the PSF stars were taken.
This procedure facilitates the scaling of the PSF stars with
respect to the PSF of SU Aur. Six sky images were also taken
before and after the observations of all the sources. These sky
frames were combined to obtain sky frames for sky subtraction.
Normalized twilight sky images were used as flat frames in the
J,H,K bands.

\subsection{Data Analysis \& the Procedure of Scaled PSF
Subtraction} Data analysis was performed using various IRAF tasks.
All images went through a standard pipeline of sky subtraction and
flat field correction before the specialized tasks of PSF matching
and subtraction. Figure 1 shows J band image of SU Aur before the
psf subtraction and a scaled and azimuthally smoothed (in sectors
of 15 degrees) mean image of PSF star. Figure 2 shows a radial 
plot (azimuthally averaged over 360 degrees) of the mean PSF star 
in Figure 1.

Since SU Aur is known to possess a large extended nebulosity
(Grady et al. 2002, Woodgate et al. 2003, Nakajima \& Golimowski
1995), the PSF matching of SU Aur with that of the PSF stars is
not a straight forward task. The usual procedure of matching the
wings of the PSF of the object (SU Aur) with that of the PSF stars
was not possible because of the presence of extended emission from
SU Aur. Therefore, an alternate procedure was adopted: 1) To get
rid of the high frequency noise and to generate a symmetric smooth
PSF function, each PSF image was azimuthally averaged over sectors
of 15 degrees. 2) A comparison of the relative brightness of SU Aur
with that of the PSF stars using their apparent magnitudes in the
respective bands was made. 3) The intensity of the PSF star images
were scaled to SU Aur. For instance, BD +40 248 (PSF1) is 1.2 mag
or 3 times brighter than SU Aur in the H band, so the intensity
from a 20 second exposure frame of PSF1 star should match that of
60 second exposure of SU Aur.

The third step was repeated for all exposures times of the PSF
stars by multiplying with suitable factors. Thus, the brightness
of both the PSF stars (PSF1 \& PSF2) were scaled to that of SU Aur
and were cross checked by subtracting one scaled PSF star from the
other until minimal intensity residuals were found in the scaled
PSF1-PSF2 images. The residual patterns of PSF1-PSF2 subtraction
are shown in Figure 3. Due to the difference in spectral type of stars
(in the present case PSF1 is G2 and PSF2 is G5) the difference in
slope of the continuum across the filter bandwidth leaves a
diffraction pattern as residuals. For example the work of Krist et
al. (1998) on PSF subtraction of one normal star from the other
using the HST data has clearly demonstrated these effects.

At the end of the third step we obtained two scaled PSF star
images which are similar. The final PSF star image was constructed
by taking mean of the two scaled PSF star images. We note here
that the x-y positions of the PSF image were matched with that of
SU Aur up to a one-fifth $(1/5)$ of a pixel for good subtraction.
Figure 4 shows PSF subtracted images of SU Aur using the same
contrast level as that of Figure 3.

The HST-STIS data (0.2 to 1.1 micron, Grady et al. 2002) in Figure
4 was downloaded from the HST archive, analyzed by the standard
STSDAS-IRAF pipeline routines and then over-plotted with the
contour levels of the J band intensity levels from 15.0 mag to 9.0
mag at an interval of 0.5 mag. We note that we did not perform PSF
subtraction on the STIS image since the STIS image is only used
for comparison of brighter regions between the optical and
near-infrared nebulosity.

Figure 5 shows mean surface brightness of the disk with respect to
the distance from the star in the J,H and K bands. The mean
surface brightness was calculated by summing up the disk
brightness in the azimuthal direction between PAs 210 to 310
degrees and dividing it by the total number of spatial resolution
element (in this case 0.3$\times$0.3 arcsecs, 4$\times$4 pixel
elements). It is clear that the procedure of PSF subtracting
images taken with the Mt.Wilson AO system can be used to detect
faint extended emission up to contrast levels of $10^{-4}$ per
resolution element (which is 0$\farcs$3$\times$0$\farcs$3) beyond
1 arcsec from the central bright star. The crosses with error bars
show the intensity of the extended nebulosity and the lines with
error bars show the residuals of PSF1 - PSF2. The accuracy of the
photometry is measured to be 0.2 mag. The lines in the J,H,K bands
therefore show instrument detection limits. In the present work we
will only consider the extended structures of SU Aur which are
beyond 1 arcsec.

\section{Results}
As evident from Figures 4 and 5 the brightness varies
between 12 to 15 mag/resolution element. The brightness of the
circumstellar nebulosity scales with the distance from the star as
$1/r^{1.2}$ to $1/r^2$ between 1 to 4 arcsec (where r is the
distance from the star in arcsecs) and maximum extent up to
3$\farcs$5 in the J band and less than 2$\farcs$5 in the K band.
The size of the circumstellar nebulosity remains intermediate in
the H band. In the J band, nebulosity is prominent from PAs 210 to
300 degrees, and faintly visible at PAs from 50 to 115 degrees.

The geometry of the extended nebulosity is more intriguing when
compared with the STIS image (Figure 4). Although the overlapping
contours of the J band image intensity show regions bright in the
STIS image, it also shows a region which is between PAs 210 to 270
degrees. In H and K bands only the latter region is seen clearly
and will be referred to as `IR nebulosity' hence forth.

Grady et al. (2002) attributed the radially streaming filamentary
structure (seen in the STIS image) at PA = 295 to 300 degrees as
the blue jet emerging from the star. It is worth mentioning here
that ground based coronagraphic images in R and I bands using
adaptive optics also show similar features (Nakajima \&
Golimowski, 1995). These coronagraph images show both the radially
outward streaming nebulosity seen in the STIS image as well as the
region between PAs 200 to 250 degrees in our near-IR images. Their
work shows brightness of about 17 and 16.5 mag/arcsec$^2$ in R and
I bands respectively at a distance of 3-4 arcsec from the star
between PA=200 to 250 degrees and about 14.5 mag/arcsec$^2$ (in R
and I) between PA= 260 to 300 degrees.

\section{Discussion}
SU Aur is known to possess a large accretion disk of size up to
400 AU and an outer envelope (Akeson et al. 2002, hereafter ACBC,
Oliveira et al. 2000). Both ground based and HST/STIS
coronagraphic images of SU Aur (Nakajima \& Golimowski 1995, Grady
et al. 2002) in the optical wavelengths have shown spatially
resolved outflows and a reflection nebula mainly due to scattered
star light. In the near-IR, however, the PSF subtracted J,H,K
images of SU Aur (present work) reveal an extended nebulosity at
angles (PA = 210 to 270) where there is little optical emission
(Figure 4, also Nakajima \& Golimowski 1995). The region where the
strongest optical nebulosity is seen is between PAs = 295 to 310
(Grady et al. 2002). Therefore, the regions probed in the near-IR are
different from those in the optical wavelengths. We discuss below
about the IR nebulosity.

There can be two explanations for the IR nebulosity: a) the region
of IR nebulosity is the opening of a cavity in the parent
molecular cloud seen almost edge on and formed by the stellar
outflow (Stapelfeldt et al.1998); b) it is the part of the
circumstellar disk itself seen at high inclination angles ($\ge$
65 degrees). In either case the region may suffer greater
extinction in the optical wavelengths (see Cotera et al. 2001). We
present the likelihoods of both the scenarios below with emphasis
on the second one (that the IR nebulosity is the part of the disk
seen in the scattered light of the central star). We also present
a geometrical model based on the disk assumption.

The IR nebulosity can be a cavity formed by the bipolar outflows
from the star (Stapelfeldt et al. 1998). Scattered light model
calculations on T-Tau and IRAS 04016+2610 by Wood et al.(2001)
have shown that cavities formed by multiple stellar outflows and
misaligned circumstellar disks can be seen in the scattered light
of the central star up to a radii of 500AU. The observed IR
nebulosity in SU Aur could be a cavity formed by the blue jet.
However, a very large opening angle of greater than 70 degrees for
the blue jet (Grady et al. 2002) would be necessary to explain the
PA of the IR nebulosity.

The IR nebulosity could be the part of the circumstellar disk seen
almost edge-on. The geometrical model describing the orientation
of the disk with respect to the observer is described by ACBC.
ACBC calculated the most probable physical parameters of the disk
like the position angle PA, angle of inclination and extent of the
disk from K band and mm interferometry and SED curve fittings.
They found that the disk outer radius could be 50 to 400 AU
(0$\farcs$3 to 2$\farcs$6) with an angle of inclination of about
62 degrees and a position angle of 130 degrees. ACBC state that
their data set on SU Aur is not sufficient to constrain the model
parameters and the quoted disk parameters by ACBC are the most
probable ones from the model fittings. However, according to STIS
coronagraphic observations (Grady et al. 2002) the blue jet is at
PA = 295 degrees and a probable red jet is at PA = 114 degrees.

We propose a schematic 3-dimensional geometrical model for SU Aur
based on our near-IR observations and the STIS observations (Grady
et al. 2002). The model (not to scale) is shown in Figure 6. The
disk is seen close to edge-on (angle of inclination = 65 degrees
or greater, from Grady et al. 2002 and ACBC) and at an azimuthal
angle of about 180 degrees (see Figure 6). Thus one side of the
disk surface faces towards the observer along with one of the jets
(the blue jet). The part of the disk surface facing towards the
observer could be seen as the IR nebulosity and the remaining part
of the disk could be obscured due to high optical extinction
(Cotera et al. 2001). According to these authors scatterd star
light through the disk can suffer from very high extinction of
Av=80 or more (also see Wood et al. 2001).

Further, if the IR nebulosity is part of the optically thick disk
seen almost edge on then we can compare the total flux of the IR
nebulosity with that of the model of ACBC. Such a comparison will
also indicate if the IR nebulosity is due to scattered light from
the star. Figure 7 of ACBC shows the flux (in Jansky) of the
theoretical model of the disk that was used to fit the SED. We
estimate an upper limit for the total flux (in Jansky) of the IR
nebulosity in each band, assuming an area covered by an annular
ring of radii 1 and 2.6 arcsecs and compared these values with the
model of ACBC. These are listed below as (values from present
work)/(model values of ACBC) in J, H, K respectively:
(0.8)/(0.10), (0.2)/(0.5), and (0.11)/(1.1).

If the model values are reliable then it is possible that in the J
band the observed flux of the IR nebulosity is dominated by the
star scattered light from the disk and the stellar jet (as in the
optical wavelengths of the STIS image) rather than the disk flux
itself. The comparison between the H band fluxes are the closest,
which may be a coincidence. It is most likely that the total flux
in the H band is dominated by the dust scattering. In the K band
we find that the observed disk flux and the model values differ by
an order of an magnitude. The scattering in the K band is less
compared to the J band and optical wavelengths and is expected to
be dominated by the disk's thermal emission. ACBC predicts the
disk temperature to be 55 to 150K at 10 AU and is expected to fall
at larger radii. Since the present work is sensitive only to
distances greater than 1 arcsec (160 AU), we are unable to see the
thermal component from the disk and measure only the scattering
component.

\section{Conclusions}
We have presented here PSF subtracted high spatial resolution images 
(0$\farcs$30 arcsecs) of the circumstellar region 
of SU Aurigae (between 1 to 4 arcsecs) in the J,H and K bands. These 
images show a distinct region bright in the near-IR which we call the 
IR nebulosity. The IR nebulosity is prominent between position angles 
210 degrees to 270 degrees and extends up to 3.5 arcsecs in the J band 
and 2.5 arcsecs in the K band. We present two scenarios about the nature 
of the IR nebulosity: a) that it could be a cavity formed by the stellar 
outflows or b) that it is a part of the circumstellar disk observed at high 
inclination angles ($\ge$65 degrees). We favor the later case more and we 
present a schematic 3-dimensional geometrical model describing the 
orientation of the disk and jet of SU Aur with respect to the observer. 
However, more observations are necessary with a large telescope equipped 
with AO and perhaps with coronagraphs and polarimetry to resolve spatial 
structures close to the star (down to 0$\farcs$1 arcsecs) for detailed 
modeling.

\acknowledgments We thank the Mt. Wilson staff for their help
during the AO observations, and Mr. Junfeng Wang for assisting the
observations. We also acknowledge Dr. Lee Hartmann and Dr. Nuria
Calvet for stimulating discussions. We thank the anonymous referee
for his/her comments which significantly improved the quality of
the paper. We also thank Julian van Eyken and John Debes for proofreading 
the manuscript carefully. This work is supported by National Science
Foundation with grant AST-0138235, and NASA grants NAG5-12115 and
NAG5-11427. This publication makes use of data products from the
Two Micron All Sky Survey, which is a joint project of the
University of Massachusetts and the Infrared Processing and
Analysis Center/California Institute of Technology, funded by the
National Aeronautics and Space Administration and the National
Science Foundation. HST-STIS data presented in this paper were obtained 
from the Multimission Archive at the Space Telescope Science Institute 
(MAST). STScI is operated by the Association of Universities for 
Research in Astronomy, Inc., under NASA contract NAS5-26555. Support 
for MAST for non-HST data is provided by the NASA Office of Space 
Science via grant NAG5-7584 and by other grants and contracts.

\clearpage

\begin{figure*}
\plotone{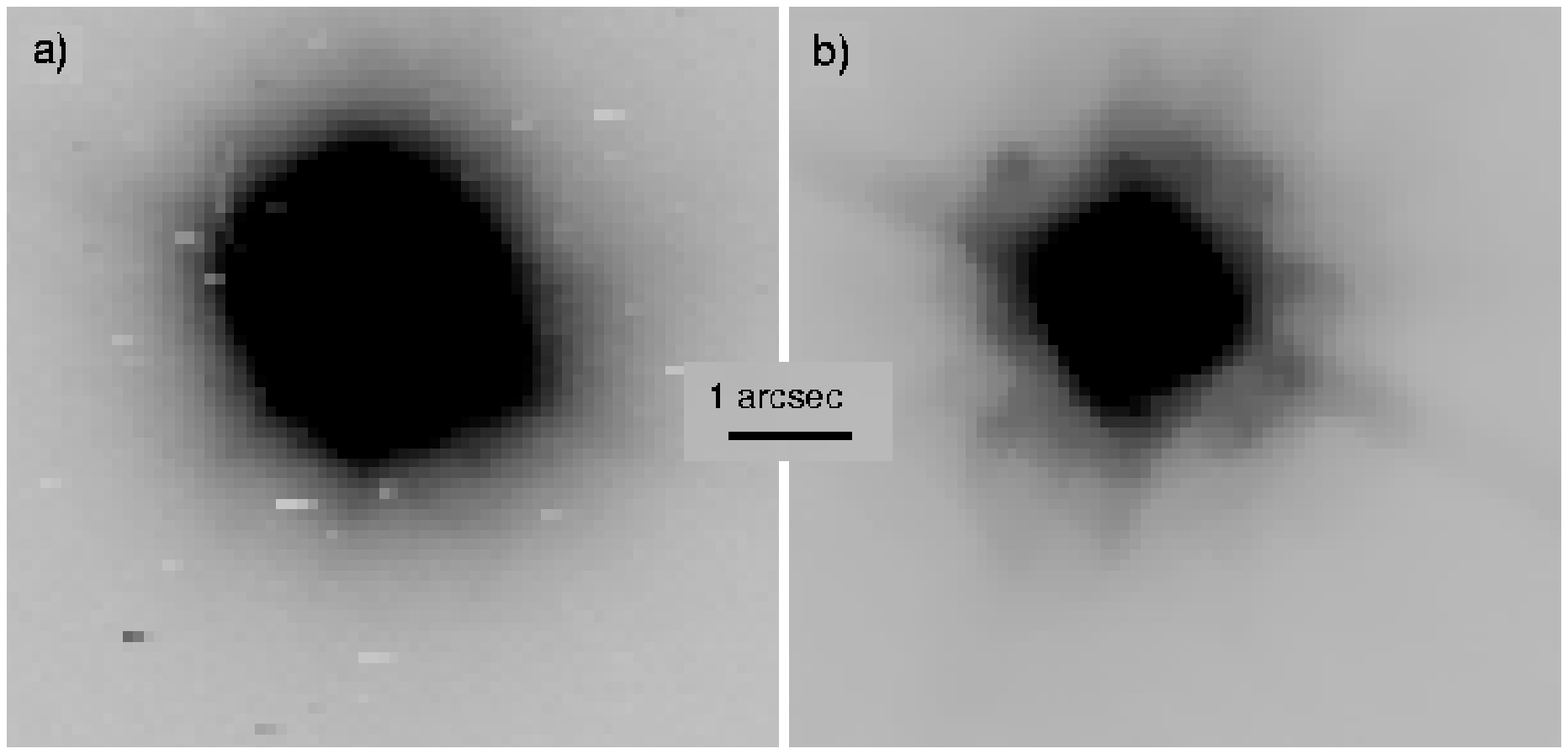} 
\caption{a) J band image of SU Aur before the
PSF subtraction. b) A scaled and azimuthally smoothed (in sectors
of 15 degrees) mean PSF star image in the J band. North is top and
East is to the left. \label{fig1}}
\end{figure*}

\begin{figure*}
\plotone{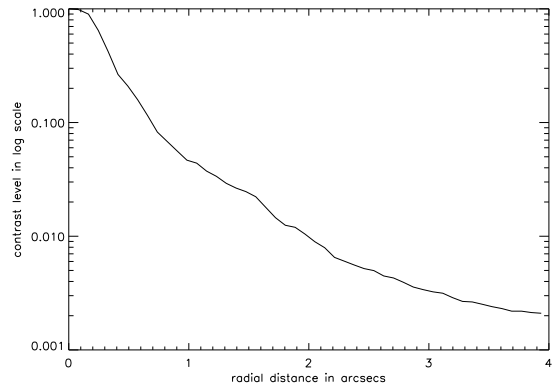} 
\caption{A radial plot (azimuthally averaged
over 360 degrees) of the mean PSF star in Figure 1 showing the
contrast levels in log scale versus radial distance in arcsecs.
\label{fig2}}
\end{figure*}

\begin{figure*}
\plotone{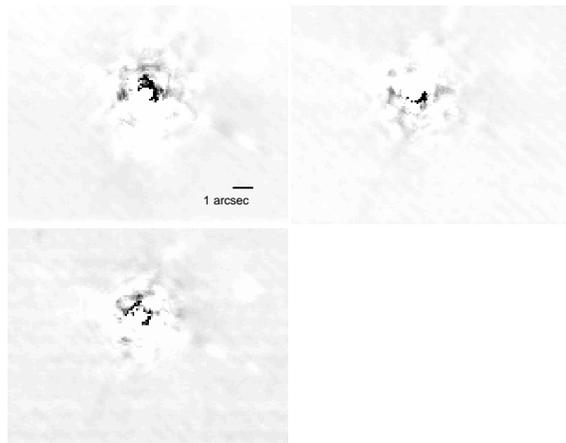} 
\caption{PSF1 (BD+40 248) subtracted from
PSF2 (Gl46.1) J(top left), H(top right), and K(bottom left) band
images. See text for details. North is up and east is to the left.
The grey scale levels are same as those in the J, H and K images
of SU Aur in Figure 4.
 \label{fig3}}

\end{figure*}
\begin{figure*}
\plotone{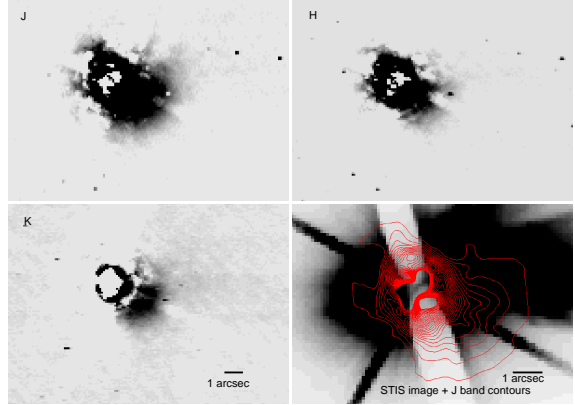} 
\caption{PSF subtracted J,H,K band images
of SU Aur showing the associated circumstellar nebulosity. See
text for details. The fourth image (bottom right) is the STIS
coronagraphic image (0.2-1.1microns) from the HST archives (also
Grady et al. 2002), and over plotted are the contours from the J
band image with contour levels from 15.0 to 9.0mag/resolution
element (see figure 5) with an interval of 0.5 mag. North is up
and east is to the left. The grey scale levels are same in the
J,H,K, and STIS images. 
Also note that the position of the diffraction spikes is along the
ridges at about PA=250 degrees seen across the nebulosity.
\label{fig4}}
\end{figure*}

\begin{figure*}
\plotone{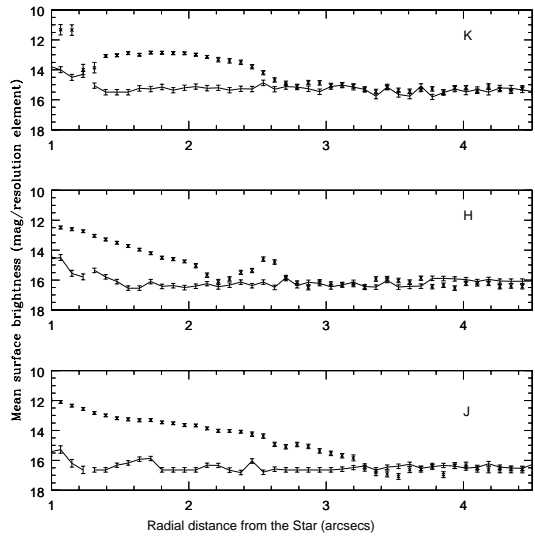} 
\caption{Mean surface brightness
(mag/resolution element, averaged azimutally between PAs 210 to
310 degrees) of the disk with respect to the projected distance
(arcsec) from the central star in J,H and K bands. Here one
resolution element is 0.3$\times$0.3 arcsec$^2$. The crosses with
error bars show the mean brightness of the disk while the simple
lines with error bars show the intensity level from the PSF1 -
PSF2 subtracted images.\label{fig5}}
\end{figure*}

\begin{figure*}
\plotone{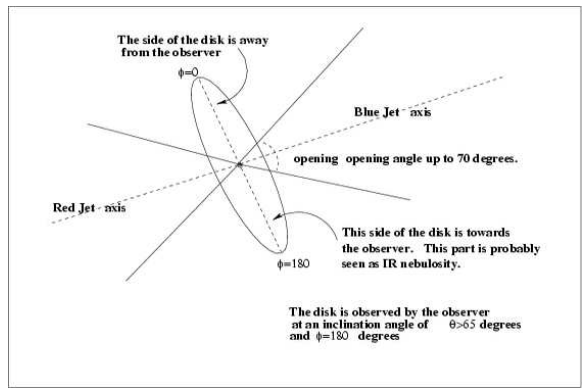} 
\caption{Schematic 3-dimensional
geometrical model (not to scale) of SU Aur based on STIS and
present NIR observations. The observer is along the direction of
$\phi$=180 degrees with the observer being able to view the face
of part of the inclined disk. This region of the disk may be seen
as the IR nebulosity.\label{fig6}}
\end{figure*}


\begin{thebibliography}{}
\bibitem[Akeson et al. (2002)]{}Akeson, R.L. et al. 2002, \apj, 566,
1124 (ACBC)
\bibitem[Cotera et al. (2001)]{}Cotera, A.S. et al. 2001, \apj, 556, 958
\bibitem[de Warf et al. (1998)]{}de Warf, L.E. et al. 1998, IAU Information Bulletin on variable
stars, No. 4551
\bibitem[de Warf et al. (2003)]{}de Warf, L.E. et al. 2003, \apj,
590, 357
\bibitem[Ge et al. (2003)]{}Ge, J., Chakraborty, A., Debes, J.H.,
Ren, D., \& Friedman, J. 2003, Proc. SPIE 4841,
1503
\bibitem[Grady et al. (2002)]{}Grady, C. et al. 2002, AAS, 199, 60.15
\bibitem[Krist et al. (1998)]{}Krist, J.E., et al. 1998, \pasp, 110, 1046
\bibitem[Nakajima \& Golimowski (1995)]{}Nakajima, T., \&
Golimowski, D.A. 1995, \aj, 109, 1181
\bibitem[Nandalin et al. (2000)]{}Nandalin, I. et al. 2000, AU Information Bulletin on variable
stars, No. 4987
\bibitem[Oliveira et al. (2000)]{}Oliveira, J.M., et al. 2000, in Disks, Planetesimals and Planets,
ASP Conference Series, ed: F. Garzn, C. Eiroa, D. Winter, and T.J. Mahoney, also astroph/0004377
\bibitem[Petrov et al. (1996)]{}Petrov, P.P. et al. 1996, \aap, 314, 821
\bibitem[Shelton et al. (1995)]{}Shelton et al. 1995, proc. SPIE,
2534, 72
\bibitem[Stapelfeldt et al. (1998)]{}Stapelfeldt, K.R. et al.
1998, \apj, 502, L65
\bibitem[Wood et al. (2001)]{}Wood, K. et al. 2001, \apj, 561, 299
\bibitem[Woodgate et al. (2003)]{}Woodgate, B. et al. 2003, proc.
SPIE, 4860, 10
\end{thebibliography}
\end{document}